\documentclass[letterpaper,12pt,notoc]{JHEP3}
\def\pd{\partial}
\def\mc{\mathcal}

\usepackage{graphicx}
\usepackage{amsmath}
\usepackage{amssymb}

\preprint{ \hbox{}\hfill arXiv: 1404.0183}

\title{RG flows in 6D N=(1,0) SCFT from SO(4) half-maximal 7D
gauged supergravity}
\author{Parinya Karndumri\\
String Theory and Supergravity Group, Department
of Physics, Faculty of Science, Chulalongkorn University, 254 Phayathai Road, Pathumwan, Bangkok 10330, Thailand\\
E-mail: \email{parinya.ka@hotmail.com}}

\abstract{We study $N=2$ seven-dimensional gauged supergravity
coupled to three vector multiplets with $SO(4)$ gauge group. The
resulting gauged supergravity contains 10 scalars consisting of the
dilaton and 9 vector multiplet scalars parametrized by
$SO(3,3)/SO(3)\times SO(3)$ coset manifold. The maximally
supersymmetric $AdS_7$ vacuum with unbroken $SO(4)$ symmetry is
identified with a $(1,0)$ SCFT in six dimensions. We find one new
supersymmetric $AdS_7$ critical point preserving
$SO(3)_{\textrm{diag}}\subset SO(3)\times SO(3)\sim SO(4)$ and study
a holographic RG flow interpolating between the $SO(4)$ and the new
$SO(3)$ supersymmetric critical points. The RG flow is driven by a
vacuum expectation value of a dimension-four operator and describes
a deformation of the UV $(1,0)$ SCFT to another supersymmetric fixed
point in the IR. In addition, a number of non-supersymmetric
critical points are identified, and some of them are stable with all
scalar masses above the BF bound. RG flows to non-conformal
$N=(1,0)$ Super Yang-Mills with $SO(2)\times SO(2)$ and $SO(2)$
symmetries are also investigated. Some of these flows have
physically acceptable IR singularities since the scalar potential is
bounded above. These provide physical RG flows from $(1,0)$ SCFT to
non-conformal field theories in six dimensions.}

\begin{document}
\section{Introduction}
The AdS/CFT correspondence has attracted a lot of attention during
the past twenty years. The original proposal in \cite{maldacena}
discussed many examples in various dimensions. These examples
included the duality between M-theory on $AdS_7\times S^4$ and
$(2,0)$ superconformal field theory (SCFT) in six dimensions. The
$AdS_7\times S^4$ geometry can arise from the near horizon limit of
M5-brane. In term of $N=4$ seven-dimensional gauged supergravity
with $SO(5)$ gauge group, the $AdS_7$ geometry corresponds to the
maximally supersymmetric vacuum of the gauged supergravity, see for
example \cite{Nishimura_AdS7CFT6}.
\\
\indent In this paper, we will explore AdS$_7$/CFT$_6$
correspondence with sixteen supercharges. The dual SCFT to the
$AdS_7$ background in this case would be $(1,0)$ six-dimensional
SCFT. Six-dimensional gauge theories with $N=(1,0)$ supersymmetry
are interesting in many aspects. In \cite{Seiberg_6D_fixed_points},
it has been shown that the theories admit non-trivial RG fixed
points. Examples of these field theories also arise in string theory
\cite{Seiberg_Witten_6Dstring}, see also a review in
\cite{Seiberg_16Q_note}. After the AdS/CFT correspondence, a
supergravity dual of a $(1,0)$ field theory with $E_8$ global
symmetry has been proposed in \cite{Berkooz_6D_dual}. The dual
gravity background has been identified with the orbifolds of
$AdS_7\times S^4$ geometry in M-theory. The operator spectrum of the
$(1,0)$ six-dimensional SCFT has been matched with the Kaluza-Klein
spectrum in \cite{AdS7_orbifold1, AdS7_orbifold2}.
\\
\indent Like in lower dimensions, it is more convenient to study
AdS$_{d+1}$/CFT$_d$ correspondence in the framework of
$(d+1)$-dimensional gauged supergravity. A consistent reduction
ansatz can eventually be used to uplift the lower dimensional
results to string/M theory in ten or eleven dimensions. A suitable
framework in the holographic study of the above $(1,0)$ field
theories is the half-maximal gauged supergravity in seven dimensions
coupled to $n$ vector multiplets. The supergravity theory has $N=2$
or sixteen supercharges in exact agreement with the number of
supercharges in six-dimensional $(1,0)$ superconformal symmetry.
This has been proposed long time ago in \cite{Ferrara_AdS7CFT6}.
With the pure gauged supergravity and critical points found in
\cite{Pure_N2_7D1} and \cite{Pure_N2_7D2}, holographic RG flows to a
non-supersymmetric IR fixed point and to a non-conformal $(1,0)$
gauge theory have been studied in \cite{Ahn_7D_flow} and
\cite{non_SUSY7Dflow}.
\\
\indent Pure $N=2$ gauged supergravity in seven dimensions admit
only two $AdS_7$ vacua with one being maximally supersymmetric and
the other one being stable non-supersymmetric. To obtain more
$AdS_7$ critical points, matter coupled supergravity theory is
needed. This has been constructed in \cite{Eric_N2_7D} but without
the topological mass term for the 3-form field which is a dual of
the 2-form field in the supergravity multipet. Without this term,
the scalar potential of the matter coupled gauged supergravity does
not admit any critical point but a domain wall as can be verified by
looking at the scalar potential explicitly given in
\cite{Eric_N2_7D}. Although mistakenly claimed in \cite{Park_7D}
that the topological mass term is not possible, the theory indeed
admits this term as shown in \cite{Eric_N2_7Dmassive} in which the
full Lagrangian and supersymmetry transformations of this massive
gauged supergravity have been given. This provides the starting
point for the present work.
\\
\indent In this paper, we are interested in the gauged supergravity
with $SO(4)$ gauge group. This requires three vector multiplets
since six gauge fields are needed in order to implement the $SO(4)$
gauging. The theory can be obtained from a truncation of the maximal
$N=4$ gauged supergravity \cite{Salam_7DN2}. In addition to the
dilaton, there are extra nine scalars from the vector multipets
parametrized by $SO(3,3)/SO(3)\times SO(3)\sim
SL(4,\mathbb{R})/SO(4)$ coset manifold. We will explore the scalar
potential of this theory in the presence of topological mass term
and identify some of its critical points. The critical points will
correspond to new IR fixed point of the $(1,0)$ SCFT identified with
the maximally supersymmetric critical point with $SO(4)$ symmetry.
We will also study RG flows between these critical points as well as
RG flows to non-conformal field theories.
\\
\indent The paper is organized as follow. We briefly review the
matter coupled gauged supergravity in seven dimensions and give
relevant formulae which will be used throughout the paper in section
\ref{N2_7D_SUGRA}. Some critical points of seven-dimensional gauged
supergravity with $SO(4)$ gauge group are explored in section
\ref{critical_point}. A number of supersymmetric and
non-supersymmetric critical points and the corresponding scalar
masses will also be given in this section. In section \ref{flow}, we
study supersymmetric deformations of the UV $N=(1,0)$ SCFT to a new
superconformal fixed point in the IR and to non-conformal SYM in six
dimensions. Both types of the solutions can be analytically
obtained. The paper is closed with some conclusions and comments on
the results in section \ref{conclusion}.

\section{$N=2$, $SO(4)$ gauged supergravity in seven dimensions}\label{N2_7D_SUGRA}
We begin with a description of $N=2$ gauged supergravity coupled to
$n$ vector multiplets. All notations are the same as those of
\cite{Eric_N2_7Dmassive}. The gravity multiplet in seven-dimensional
$N=2$ supersymmetry contains the following field content
\begin{equation}
\textrm{gravity multiplet}:\qquad (e^m_\mu, \psi^A_\mu,
A^i_\mu,\chi^A,B_{\mu\nu},\sigma).
\end{equation}
A vector multiplet has the field content $(A_\mu,\lambda^A,\phi^i)$.
Indices $A,B$ label the doublet of the $USp(2)_R\sim SU(2)_R$
R-symmetry. Curved and flat space-time indices are denoted by
$\mu,\nu,\ldots$ and $m,n,\ldots$, respectively. $B_{\mu\nu}$ and
$\sigma$ are a two-form and the dilaton fields. For supergravity
theory coupled to $n$ vector multiplets, there are $n$ copies of
$(A_\mu,\lambda^A,\phi^i)^r$ labeled by an index $r=1,\ldots, n$,
and indices $i,j=1,2,3$ label triplets of $SU(2)_R$. The $3n$
scalars $\phi^{ir}$ are parametrized by $SO(3,n)/SO(3)\times SO(n)$
coset manifold. The corresponding coset representative will be
denoted by
\begin{equation}
L=(L_I^{\phantom{s}i},L_I^{\phantom{s}r}), \qquad I=1,\ldots, n+3\,
.
\end{equation}
The inverse of $L$ is given by
$L^{-1}=(L^I_{\phantom{s}i},L^I_{\phantom{s}r})$ where
$L^I_{\phantom{s}i}=\eta^{IJ}L_{Ji}$ and
$L^I_{\phantom{s}r}=\eta^{IJ}L_{Jr}$. Indices $i,j$ and $r,s$ are
raised and lowered by $\delta_{ij}$ and $\delta_{rs}$, respectively
while the full $SO(3,n)$ indices $I,J$ are raised and lowered by
$\eta_{IJ}=\textrm{diag}(---++\ldots +)$. There are some relations
involving components of $L$ and are given by
\begin{eqnarray}
\eta_{IJ}&=&-L_I^{\phantom{a}i}L^i_{\phantom{s}J}+L_I^{\phantom{a}r}L_J^{\phantom{s}r},\qquad
L^i_{\phantom{I}}=L_{Ii},\nonumber \\
L^i_{\phantom{a}I}L^I_{\phantom{s}j}&=&-\delta^i_j,\qquad
L^i_{\phantom{a}I}L^{Ij}=-\delta^{ij}\, .
\end{eqnarray}
\indent Gaugings are implemented by promoting a global symmetry
$\tilde{G}\subset SO(3,n)$ to a gauge symmetry. Consistency of the
gauging imposes a condition on the $\tilde{G}$ structure constants
$f_{IJ}^{\phantom{sas}K}$
\begin{equation}
f_{IK}^{\phantom{sas}L}\eta_{LJ}+f_{JK}^{\phantom{sas}L}\eta_{LI}=0
\end{equation}
meaning that $\eta_{IJ}$ is invariant under the adjoint action of
$\tilde{G}$. General semisimple gauge groups take the form of
$\tilde{G}\sim G_0\times H\subset SO(3,n)$ with $G_0$ being one of
the six possibilities: $SO(3)$, $SO(3,1)$, $SL(3,\mathbb{R})$,
$SO(2,1)$, $SO(2,2)$ and $SO(2,2)\times SO(2,1)$ and $H$ being
compact with $\textrm{dim}H\leq (n+3-\textrm{dim}G_0)$.
\\
\indent In this paper, we are interested in the $SO(4)$ gauged
supergravity corresponding to $G_0=SO(3)$ and $H=SO(3)$. To obtain
$AdS_7$ vacua, we need to consider the gauged supergravity with a
topological mass term for a 3-form potential. The 3-form field is a
dual of the 2-form $B_{\mu\nu}$. With all modifications to the
Lagrangian and supersymmetry transformations as given in
\cite{Eric_N2_7Dmassive}, the bosonic Lagrangian involving only
scalars and the metric can be written as
\begin{equation}
e^{-1}\mc{L}=\frac{1}{2}R-\frac{5}{8}\pd_\mu \sigma \pd^\mu
\sigma-\frac{1}{2}P^{\mu ir}P_{\mu ir}-V
\end{equation}
where the scalar potential is given by
\begin{equation}
V=\frac{1}{4}e^{-\sigma}\left(C^{ir}C_{ir}-\frac{1}{9}C^2\right)+16h^2e^{4\sigma}
-\frac{4\sqrt{2}}{3}he^{\frac{3\sigma}{2}}C\,.
\end{equation}
The constant $h$ characterizes the topological mass term. The
quantities appearing in the above equations are defined by
\begin{eqnarray}
P_\mu^{ir}&=&L^{Ir}\left(\delta^K_I\pd_\mu+f_{IJ}^{\phantom{sad}K}A_\mu^J\right)L^i_{\phantom{s}K},
\qquad
C_{rsi}f_{IJ}^{\phantom{sad}K}L^I_{\phantom{s}r}L^J_{\phantom{s}s}L_{Ki},\nonumber
\\
C_{ir}&=&\frac{1}{\sqrt{2}}f_{IJ}^{\phantom{sad}K}L^I_{\phantom{s}j}L^J_{\phantom{s}k}L_{Kr}\epsilon^{ijk},
\qquad
C=-\frac{1}{\sqrt{2}}f_{IJ}^{\phantom{sad}K}L^I_{\phantom{s}i}L^J_{\phantom{s}j}L_{Kk}\epsilon^{ijk}\,
.
\end{eqnarray}
We also need fermionic supersymmetry transformations with all fields
but scalars vanishing. These are given by
\begin{eqnarray}
\delta \psi_\mu &=&2D_\mu
\epsilon-\frac{\sqrt{2}}{30}e^{-\frac{\sigma}{2}}C\gamma_\mu
\epsilon -\frac{4}{5}he^{2\sigma}\gamma_\mu \epsilon,\label{delta_psi}\\
\delta \chi &=&-\frac{1}{2}\gamma^\mu\pd_\mu \sigma
\epsilon+\frac{\sqrt{2}}{30}e^{-\frac{\sigma}{2}}C\epsilon-\frac{16}{5}e^{2\sigma}h\epsilon,\label{delta_chi}\\
\delta \lambda^r &=&-i\gamma^\mu
P^{ir}_\mu\sigma^i\epsilon-\frac{i}{\sqrt{2}}e^{-\frac{\sigma}{2}}C^{ir}\sigma^i\epsilon\label{delta_lambda}
\end{eqnarray}
where $SU(2)_R$ indices on spinors are suppressed. $\sigma^i$ are
the usual Pauli matrices.
\\
\indent In the remaining of this section, we focus on $n=3$ case
with $\tilde{G}=SO(4)\sim SO(3)\times SO(3)$. The first $SO(3)$
factor is identified with the $SU(2)_R$ R-symmetry. To give an
explicit parametrization of $SO(3,3)/SO(3)\times SO(3)$ coset, we
define thirty-six $6\times 6$ matrices
\begin{equation}
(e_{ab})_{cd}=\delta_{ac}\delta_{bd},\qquad a,b\ldots =1,\ldots 6\,
.
\end{equation}
Non-compact generators of $SO(3,3)$ are identified as
\begin{equation}
Y_{ir}=e_{i,r+3}+e_{r+3,i},\qquad r=1,\ldots, 3\, .
\end{equation}
Accordingly, $SO(3)\times SO(3)$ generators can be written as
\begin{eqnarray}
SO(3)_R &:&\qquad J_{ij}=e_{ij}-e_{ji},\nonumber \\
SO(3) &:&\qquad J_{rs}=e_{rs}-e_{sr}\, .
\end{eqnarray}
In this case, the structure constants for the gauge group are given
by
\begin{equation}
f_{IJK}=(g_1\epsilon_{ijk},g_2\epsilon_{rst})
\end{equation}
where $g_1$ and $g_2$ are coupling constants of $SO(3)_R$ and
$SO(3)$, respectively.

\section{Critical points of $N=2$, $SO(4)$ seven-dimensional gauged supergravity}\label{critical_point}
In this section, we will compute the scalar potential of the $SO(4)$
gauged supergravity and study some of its critical points. Although
complicated, it is possible to compute the scalar potential for all
of the ten scalars. However, the long expression would make any
analysis more difficult. Consequently, we will proceed by studying
the scalar potential on a subset of the ten scalars as originally
proposed in \cite{warner}. In this approach, the scalar potential is
computed on a scalar submanifold which is invariant under some
subgroup $H_0$ of the full gauge symmetry $SO(4)$. This submanifold
consists of all scalars which are singlet under the unbroken
subgroup $H_0$. All critical points found on this submanifold are
essentially critical points of the potential on the full scalar
manifold. This can be seen by expanding the full potential to first
order in scalar fluctuations which in turn contain both $H_0$
singlets and $H_0$ non-singlets. By a simple group theory argument,
the non-singlet fluctuations cannot lead to $H_0$ singlets at first
order. Their coefficients, variations of the potential with respect
to non-singlet scalars, must accordingly vanish. This proves to be
more convenient and more efficient. However, the truncation is
consistent only when all relevant $H_0$ singlet scalars are included
on the chosen submanifold. With only some of these singlets, the
consistency is not guaranteed.

\subsection{Critical points on $SO(3)_{\textrm{diag}}$ scalars}
We begin with the most simplest case namely the potential on
$SO(3)_{\textrm{diag}}\subset SO(3)\times SO(3)$ corresponding to
the non-compact generator $Y_s=Y_{11}+Y_{22}+Y_{33}$. The coset
representative is then parametrized by
\begin{equation}
L=e^{\phi Y_s}\, .
\end{equation}
The scalar potential is given by
\begin{eqnarray}
V&=&\frac{1}{32}e^{-\sigma}\left[(g_1^2+g_2^2)\left(\cosh
(6\phi)-9\cosh(2\phi)\right)-8g_1g_2\sinh^3(2\phi)\phantom{e^{\frac{1}{2}}}\right.\nonumber
\\ & &\left.+8\left[g_2^2-g_1^2+64h^2e^{5\sigma}+32e^{\frac{5\sigma}{2}}h\left(g_1\cosh^3\phi-g_2\sinh^3\phi\right)\right]
\right].
\end{eqnarray}
\indent Notice that there is no critical point when $h=0$ as
mentioned before. In this case, the $SO(4)$ supergravity admits a
half-supersymmetric domain wall as a vacuum solution. For $\phi=0$,
the above potential is the potential of pure $N=2$ gauged
supergravity with $SO(3)$ gauge group studied in \cite{Pure_N2_7D1}
and \cite{Pure_N2_7D2}. There are two critical points in the pure
gauged supergravity. One of them preserves all of the supersymmetry
while the other completely breaks supersymmetry. In our conventions,
they are given by
\begin{equation}
\sigma =\frac{2}{5}\ln \left[-\frac{g_1}{16h}\right]\qquad
\textrm{and}\qquad \sigma =\frac{2}{5}\ln
\left[-\frac{g_1}{8h}\right].
\end{equation}
It can be readily verified by using supersymmetry transformations of
$\psi_\mu$, $\chi$ and $\lambda^r$ that the first one is
supersymmetric. We can bring the supersymmetric point to $\sigma=0$
by choosing $g_1=-16h$ and find that the two critical points are now
given by
\begin{eqnarray}
\sigma &=&0,\qquad V_0=-240h^2\nonumber \\
\textrm{and}\qquad \sigma &=&\frac{2}{5}\ln 2,\qquad V_0=-160
(2^{\frac{3}{5}})h^2
\end{eqnarray}
where $V_0$ denotes the value of the cosmological constant.
\\
\indent Although non-supersymmetric, the second critical point has
been shown to be stable in \cite{Pure_N2_7D2}. In the presence of
matter scalars, this is however not the case. This can be seen from
the scalar masses given below.
\\
\begin{center}
\begin{tabular}{|c|c|}
  \hline
  $SO(3)\times SO(3)$ & $m^2L^2$ \\ \hline
  $(\mathbf{1},\mathbf{1})$ & $12$ \\
  $(\mathbf{3},\mathbf{3})$ & $-12$ \\
  \hline
\end{tabular}
\end{center}
The $AdS_7$ radius $L$ in our conventions is given by
$L=\sqrt{\frac{-15}{V_0}}=\frac{1}{4h}$. The
$(\mathbf{1},\mathbf{1})$ scalar correspond to $\sigma$, and
$(\mathbf{3},\mathbf{3})$ is the nine scalars in
$SO(3,3)/SO(3)\times SO(3)$. The BF bound in seven dimensions is
$m^2L^2\geq -9$. Therefore, the non-supersymmetric critical point of
pure gauged supergravity is unstable in the matter coupled theory.
This is very similar to the six-dimensional $N=(1,1)$ gauged
supergravity pointed out in \cite{F4_flow}.
\\
\indent Scalar masses at the supersymmetric point are given in the
table below.
\\
\begin{center}
\begin{tabular}{|c|c|}
  \hline
  $SO(3)\times SO(3)$ & $m^2L^2$ \\ \hline
  $(\mathbf{1},\mathbf{1})$ & $-8$ \\
  $(\mathbf{3},\mathbf{3})$ & $-8$ \\
  \hline
\end{tabular}
\end{center}
In the dual $(1,0)$ SCFT, these scalars correspond to dimension-4
operators via the relation $m^2L^2=\Delta(\Delta-6)$.
\\
\indent There is one non-trivial supersymmetric point at
\begin{eqnarray}
\sigma &=&-\frac{1}{5}\ln
\left[\frac{g_2^2-256h^2}{g_2^2}\right],\qquad
\phi=\frac{1}{2}\ln\left[\frac{g_2-16h}{g_2+16h}\right],\nonumber
\\
V_0&=&-\frac{240g_2^{\frac{8}{5}}h^2}{(g_2^2-256h^2)^{\frac{4}{5}}}\,
.
\end{eqnarray}
At this point, scalar masses are computed as follow.
\\
\begin{center}
\begin{tabular}{|c|c|c|}
  \hline
  $SO(3)_{\textrm{diag}}$ & $m^2L^2$ &$\Delta$ \\ \hline
  $\mathbf{1}$ & $-8$ & $4$\\
  $\mathbf{1}$ & $40$ & $10$ \\
  $\mathbf{3}$ & $0$  & $6$ \\
  $\mathbf{5}$ & $16$ & $8$ \\
  \hline
\end{tabular}
\end{center}
In the table, we have decomposed all of the ten scalars in
representations of the $SO(3)_{\textrm{diag}}$ residual symmetry.
This can be done by the following decomposition. Under
$SO(3)_R\times SO(3)$, the nine scalars transform as
$(\mathbf{3},\mathbf{3})$. They then transform as
$\mathbf{3}\times\mathbf{3}=\mathbf{1}+\mathbf{3}+\mathbf{5}$ under
$SO(3)_{\textrm{diag}}$. Notice that the $\mathbf{3}$ scalars are
massless corresponding to Goldstone bosons of the symmetry breaking
$SO(3)\times SO(3)\rightarrow SO(3)_{\textrm{diag}}$.
\\
\indent There is one non-supersymmetric critical point given by
\begin{eqnarray}
\sigma
&=&\frac{1}{5}\ln\left[\frac{4g_2^2}{g_2^2-256h^2}\right],\qquad
\phi=\frac{1}{2}\ln
\left[\frac{g_2-16h}{g_2+16h}\right],\nonumber \\
V_0&=&-\frac{160(2^{\frac{3}{5}})g_2^{\frac{8}{5}}h^2}{(g_2^2-256h^2)^{\frac{4}{5}}}\,
.
\end{eqnarray}
This critical point is stable as can be seen from the mass spectrum
below.
\\
\begin{center}
\begin{tabular}{|c|c|c|}
  \hline
  $SO(3)_{\textrm{diag}}$ & $m^2L^2$ & $\Delta$ \\ \hline
  $\mathbf{1}$ & $12$ & $3+\sqrt{21}$\\
  $\mathbf{1}$ & $36$ & $3+3\sqrt{5}$ \\
  $\mathbf{3}$ & $0$ & $6$ \\
  $\mathbf{5}$ & $0$ & $6$ \\
  \hline
\end{tabular}
\end{center}
\indent For $g_2=g_1$, we also find another non-supersymmetric
critical point given by
\begin{equation}
\sigma = \frac{1}{10}\left[\sqrt{2}\ln 8+4\ln
(1-2^{-\sqrt{2}})\right],\qquad \phi=-\frac{1}{2}\ln 2,\qquad
V_0=-246.675h^2\, .
\end{equation}
This critical point is however unstable. Scalar masses at this point
are given below.
\\
\begin{center}
\begin{tabular}{|c|c|}
  \hline
  $SO(3)_{\textrm{diag}}$ & $m^2L^2$ \\ \hline
  $\mathbf{1}$ & $-4.278$ \\
  $\mathbf{1}$ & $16.059$ \\
  $\mathbf{3}$ & $0$ \\
  $\mathbf{5}$ & $-14.282$ \\
  \hline
\end{tabular}
\end{center}
We can see that the mass of $\mathbf{5}$ scalars violates the BF
bound.

\subsection{Critical points on scalar manifold with smaller residual symmetry}
To find other critical points, we can consider smaller residual
symmetries. Breaking $SO(3)_{\textrm{diag}}$ to
$SO(2)_{\textrm{diag}}$, we find that there are two singlets from
$SO(3,3)/SO(3)\times SO(3)$ with the coset representative
\begin{equation}
L=e^{\phi_1(Y_{11}+Y_{22})}e^{\phi_2Y_{33}}\, .
\end{equation}
This gives the scalar potential, with $g_1=-16h$,
\begin{eqnarray}
V&=&\frac{1}{8}e^{-\sigma}\left[2(g_2^2+64h^2(e^{5\sigma}-4))-2(g_2^2+256h^2)\cosh(2\phi_1)\right.\nonumber
\\
& &-64he^{\frac{5\sigma}{2}}\left(16h\cosh^2
\phi_1\cosh\phi_2+g_2\sinh^2\phi_1\sinh \phi_2\right)\nonumber \\
&
&\left.+\sinh^2(2\phi_1)\left[(g_2^2+256h^2)\cosh(2\phi_2)+32g_2h\sinh(2\phi_2)\right]
\right].\label{V_SO2D}
\end{eqnarray}
This potential does not admit any supersymmetric critical points
unless $\phi_1=\phi_2$ which is the previously found
$SO(3)_{\textrm{diag}}$ point. When $\phi_1=0$, the above scalar
submanifold preserves $SO(2)\times SO(2)$ symmetry, but there is no
critical point except for $\phi_2=0$. We are not able to obtain any
new critical points from the above potential.
\\
\indent We now move to scalar fields invariant under $SO(2)_R\subset
SO(3)_R$. There are three singlets corresponding to $Y_{11}$,
$Y_{12}$ and $Y_{13}$. Denoting the associated scalars by $\phi_i$,
$i=1,2,3$, we find a simple potential
\begin{equation}
V=-\frac{1}{2}g_1^2e^{-\sigma}+16h^2e^{4\sigma}+g_1he^{\frac{3}{2}\sigma-\phi_1-\phi_2-\phi_3}
(1+e^{2\phi_1})(1+e^{2\phi_2})(1+e^{2\phi_3})\label{V_SO2R}
\end{equation}
which does not admit any non-trivial critical points.

\section{Supersymmetric RG flows}\label{flow}
We now consider domain wall solutions interpolating between critical
points identified in the previous section. These solutions will
generally have an interpretation in terms of RG flows in the dual
field theories in six dimensions. We are mainly interested in
supersymmetric RG flows which can be obtained from solving BPS
equations coming from supersymmetry variations of fermionic fields
$\psi_\mu$, $\chi$ and $\lambda^r$. A stable non-supersymmetric
$AdS_7$ critical point also admits a well-defined dual CFT, but in
most cases, finding the corresponding flow solutions requires a
numerical analysis. Accordingly, we will not consider
non-supersymmetric flows in this paper.

\subsection{An RG flow to a supersymmetric $SO(3)$ fixed point}
There is one supersymmetric $AdS_7$ critical point with $SO(3)$
symmetry. In this subsection, we will find the domain wall solution
interpolating between this point and the trivial critical point at
$\sigma=\phi=0$.
\\
\indent Using the standard domain wall metric
\begin{equation}
ds^2=e^{2A(r)}dx^2_{1,5}+dr^2
\end{equation}
where $dx^2_{1,5}$ is the flat metric in six-dimensional space-time
and the projection condition $\gamma_r\epsilon=\epsilon$, we can
derive the following BPS equations
\begin{eqnarray}
\phi'&=&\frac{1}{8}e^{-\frac{\sigma}{2}-3\phi}(e^{4\phi}-1)\left(g_1+g_2+e^{2\phi}g_1-e^{2\phi}g_2\right),\label{eq1}\\
\sigma'&=&\frac{1}{20}\left[e^{-\frac{\sigma}{2}-3\phi}\left(g_2(e^{2\phi}-1)^3
-g_1(1+e^{2\phi})^3\right)-128he^{2\sigma}\right],\label{eq2}\\
A'&=&\frac{1}{40}e^{-\frac{\sigma}{2}-3\phi}\left[g_2(e^{2\phi}-1)^3
-g_1(1+e^{2\phi})^3\right]+\frac{4}{5}he^{2\sigma}\label{eq3}
\end{eqnarray}
where $'$ denotes $\frac{d}{dr}$. The above equations do not involve
$\delta\psi_r$ equation which will give the Killing spinor condition
on $\epsilon$ as usual. The above equations clearly admit two
critical points. To find the solution, we combine equations
\eqref{eq1} and \eqref{eq2} to
\begin{equation}
\frac{d\sigma}{d\phi}=\frac{2\left[g_2(e^{2\phi}-1)^3-g_1(1+e^{2\phi})^3-128he^{\frac{\sigma}{2}+3\phi}\right]}
{5(e^{4\phi}-1)\left(g_1+g_2+(g_1-g_2)e^{2\phi}\right)}
\end{equation}
whose solution is given by
\begin{equation}
\sigma=\frac{2}{5}\ln
\left[\frac{e^\phi\left(g_1+g_2+(g_1-g_2)e^{2\phi}\right)}{32h\left(12C_1(e^{2\phi}-1)-1\right)}\right].
\end{equation}
In order for the solution to interpolate between the two critical
points, we need to fix the integration constant to be
$C_1=\frac{(g_1-g_2)^2}{48g_1g_2}$. We then find the solution for
$\sigma$
\begin{equation}
\sigma = \frac{2}{5}\ln
\left[-\frac{g_1g_2e^\phi}{8h\left(g_1+g_2+(g_2-g_1)e^{2\phi}\right)}\right].
\end{equation}
\indent Introducing a new radial coordinate $\tilde{r}$ via
$\frac{d\tilde{r}}{dr}=e^{-\frac{\sigma}{2}}$, we can solve equation
\eqref{eq1} and find the solution for $\phi$
\begin{equation}
g_1g_2\tilde{r}=2g_1\tan^{-1}e^\phi+2\sqrt{g_2^2-g_1^2}\tanh^{-1}\left[e^\phi\sqrt{\frac{g_2-g_1}{g_2+g_1}}\right]
+g_2\ln\left[\frac{1-e^\phi}{1+e^\phi}\right]
\end{equation}
where we have neglected an additive integration constant to
$\tilde{r}$. Taking the combination \eqref{eq3}+$\frac{1}{8}\times$
\eqref{eq2} and changing the variable from $r$ to $\phi$, we find
\begin{equation}
\frac{dA}{d\phi}+\frac{1}{8}\frac{d\sigma}{d\phi}=\frac{g_2(e^{2\phi}-1)^3-g_1(1+e^{2\phi})^3}
{4(e^{4\phi}-1)\left(g_1+g_2+(g_1-g_2)e^{2\phi}\right)}\, .
\end{equation}
The solution is easily found to be
\begin{equation}
A=\frac{1}{8}\left[2\phi-\sigma-2\ln \left(2-2e^{4\phi}\right)+2\ln
\left(g_1+g_2+(g_1-g_2)e^{2\phi}\right)\right].
\end{equation}
\indent Near the UV point $\sigma\sim 0$ and $\phi\sim 0$ with
$g_1=-16h$, we find
\begin{equation}
\sigma \sim \phi\sim e^{-16hr}=e^{-\frac{4r}{L}},\qquad
L=\frac{1}{4h}
\end{equation}
since $\tilde{r}\sim r$ near $\sigma \sim 0$. The flow is then
driven by vacuum expectation values (vev) of relevant operators of
dimension $\Delta=4$. In the IR, we find that the solution behaves
as
\begin{equation}
\sigma \sim e^{-\frac{4r}{L}},\qquad \phi\sim
e^{\frac{4r}{L}},\qquad
L=\frac{(g_2^2-256h^2)^{\frac{2}{5}}}{4hg_2^{\frac{4}{5}}}\, .
\end{equation}
From this, we see that the operator dual to $\phi$ acquires an
anomalous dimension and has dimension $10$ in the IR. This is
consistent with the value of $m^2L^2$ given previously.

\subsection{RG flows to non-conformal field theories}
A supersymmetric flow to non-conformal field theory in pure gauged
supergravity has been studied in \cite{non_SUSY7Dflow}. We will
study similar solutions in the matter coupled gauged supergravity.
These solutions would be a generalization of the solution given in
\cite{non_SUSY7Dflow}.

\subsubsection{Flows to $SO(2)\times SO(2)$, $6D$ Super Yang-Mills}
We first consider $SO(2)_R$ singlets scalars. With
$\gamma_r\epsilon=\epsilon$, the BPS equations for these three
singlets, denoted by $\phi_i$, $i=1,2,3$, $\sigma$ and $A$ are given
by
\begin{eqnarray}
\phi_1'&=&\frac{1}{2}e^{-\frac{\sigma}{2}-\phi_1}g_1(e^{2\phi_1}-1),\\
\phi_2'&=&\frac{1}{2}e^{-\frac{\sigma}{2}-\phi_2}g_1(e^{2\phi_2}-1),\\
\phi_3'&=&\frac{1}{2}e^{-\frac{\sigma}{2}-\phi_3}g_1(e^{2\phi_3}-1),\\
\sigma'&=&-\frac{1}{20}g_1e^{-\frac{\sigma}{2}-\phi_1-\phi_2-\phi_3}(1+e^{2\phi_1})
(1+e^{2\phi_2})(1+e^{2\phi_3})-\frac{32}{5}he^{2\sigma},\\
A'&=&-\frac{1}{40}g_1e^{-\frac{\sigma}{2}-\phi_1-\phi_2-\phi_3}(1+e^{2\phi_1})
(1+e^{2\phi_2})(1+e^{2\phi_3})+\frac{4}{5}he^{2\sigma}\, .
\end{eqnarray}
The above equations clearly admit only one critical point at
$\phi_i=0$.
\\
\indent For $\phi_1=\phi_2=0$, the solution will preserve
$SO(2)_R\times SO(2)$ symmetry. This is easily seen to be a
consistent truncation. The solution to the above equations is given
by
\begin{eqnarray}
\phi_3 &=&\pm\ln
\left[\frac{1+e^{g_1\tilde{r}+C_1}}{1-e^{g_1\tilde{r}+C_1}}\right],\nonumber
\\
\sigma &=&\frac{2}{5}\phi_3 -\frac{2}{5}\ln
\left[-\frac{16h}{g_1}\left[4C_2\left(e^{2\phi_3}-1\right)-1\right]
\right],\nonumber
\\
A&=&\frac{1}{8}\left[2\phi_3-\sigma-2\ln (e^{2\phi_3}-1)\right]
\end{eqnarray}
where as in the previous case $\tilde{r}$ is related to $r$ via
$\frac{d\tilde{r}}{dr}=e^{-\frac{\sigma}{2}}$.
\\
\indent Near the UV point, the asymptotic behavior of $\phi_3$ and
$\sigma$ is given by
\begin{equation}
\phi_3\sim \sigma \sim e^{-16hr},\qquad A\sim 4hr\sim \frac{r}{L}\,
.
\end{equation}
\indent In the IR, we will consider $\phi_3>0$ and $\phi_3<0$,
separately. For $\phi_3>0$, there is a singularity when
$\phi_3\rightarrow \infty$ as $16h\tilde{r}\sim C_1$. With $C_2\neq
0$, we find
\begin{eqnarray}
& &\phi_3\sim -\ln (16h\tilde{r}-C_1),\qquad \sigma \sim
\frac{2}{5}\ln(16h\tilde{r}-C_1),\nonumber
\\& &
A\sim
-\frac{1}{8}(2\phi_3+\sigma)=\frac{1}{5}\ln(16h\tilde{r}-C_1)\,
.\label{first_sol}
\end{eqnarray}
As $16h\tilde{r}\sim C_1$, we find the relation between $r$ and
$\tilde{r}$ to be
$16hr-C=\frac{5}{6}(16h\tilde{r}-C_1)^{\frac{6}{5}}$ with $C$ being
another integration constant. As expected from the general DW/QFT
correspondence
\cite{DW/QFT_townsend,correlator_DW/QFT,Skenderis_DW/QFT}, the
metric in the IR takes the form of a domain wall
\begin{equation}
ds^2=(16hr-C)^{\frac{1}{3}}dx^2_{1,5}+dr^2
\end{equation}
where the multiplicative constant has been absorbed in the rescaling
of the $x^\mu$ coordinates.
\\
\indent Flows to non-conformal field theories usually encounter
singularities in the IR. As can be seen from the above metric, there
is a singularity at $16hr\sim C$. The criterion for determining
whether a given singularity is physical or not has been given in
\cite{Gubser_singularity}. The condition rules out naked time-like
singularities which are clearly unphysical. According to the
criterion of \cite{Gubser_singularity}, the IR singularity in the
solution is acceptable if the scalar potential is bounded above. One
way to understand this criterion has been given in
\cite{non_CFT_flow} for four-dimensional gauge theories. We will
follow this argument and briefly discuss the meaning of the
criterion in \cite{Gubser_singularity} in the context of
six-dimensional field theories. Near the IR singularity, scalars
$\phi_i$, assumed to be canonical ones, and the metric warped factor
$A$ behave as
\begin{equation}
\phi_i\sim B_i\ln (r-r_0),\qquad A\sim \kappa \ln (r-r_0)
\end{equation}
where we have chosen the integration constant so that the
singularity occurs at $r_0$. In the IR, the bulk action for these
scalars mainly contains the kinetic terms since the potential is
irrelevant. This is because the potential diverges logarithmically,
but the kinetic terms go like $(r-r_0)^{-2}$. According to the
AdS/CFT correspondence, the one point function or the vacuum
expectation value of operators $O_i$ dual to $\phi_i$ is given by
$\langle O_i\rangle =\frac{\delta S}{\delta \phi_i}$. Using
\begin{equation}
S=\frac{1}{2}\int d^6x dr e^{6A}\pd_r \phi_i\pd^r \phi_i,
\end{equation}
we find
\begin{equation}
\langle O_i\rangle = \frac{\delta S}{\delta \phi_i}\sim e^{6A}\pd_r
\phi_i\sim B_i (r-r_0)^{6\kappa-1}\, .
\end{equation}
We can see that $\langle O_i\rangle$ diverges for $\kappa <
\frac{1}{6}$. We then expect that solutions with $\kappa <
\frac{1}{6}$ will be excluded. In four dimensions, it has been shown
that this is related to the fact that the scalar potential becomes
unbounded above. In the present case, we will see in the solutions
given below that this is the case namely all solutions with $\kappa
< \frac{1}{6}$ have $V\rightarrow \infty$.
\\
\indent It can be checked by using the scalar potential given in
\eqref{V_SO2R} that as $16h\tilde{r}\sim C_1$, the solution in
\eqref{first_sol} gives $V\rightarrow -\infty$. The solution is then
physical and describes a supersymmetric RG flow from $(1,0)$ SCFT to
six-dimensional SYM with $SO(2)\times SO(2)$ symmetry.
\\
\indent For $C_2=0$, the solution becomes
\begin{eqnarray}
\phi_3&\sim &-\ln (16h\tilde{r}-C_1),\qquad \sigma\sim
-\frac{2}{5}\ln (16h\tilde{r}-C_1),\nonumber \\
ds^2&=&(16hr-C)^{\frac{3}{4}}dx^2_{1,5}+dr^2\, .
\end{eqnarray}
This is also physical since it leads to $V\rightarrow -\infty$.
\\
\indent For $\phi_3<0$ and $16h\tilde{r}\sim C_1$, the above
solutions give, for any values of $C_2$,
\begin{eqnarray}
& &\phi_3\sim \ln (16h\tilde{r}-C_1),\qquad \sigma \sim
\frac{2}{5}\ln(16h\tilde{r}-C_1),\nonumber \\
& &ds^2=(16hr-C)^{\frac{1}{3}}dx^2_{1,5}+dr^2
\end{eqnarray}
which give rise to $V\rightarrow -\infty$. This solution is then
physically acceptable.
\\
\indent The solution with all $\phi_i\neq 0$ turns out to be very
difficult to find although the above BPS equations suggest that
$\phi_1=\phi_2=\phi_3$. Most probably, a numerical analysis might be
needed. Therefore, we will not further investigate this case.

\subsubsection{Flows to $SO(2)$, $6D$ Super Yang-Mills}
As a final example, we consider RG flows to non-conformal theories
from $SO(2)_{\textrm{diag}}$ singlet scalars corresponding to
$Y_{11}+Y_{22}$ and $Y_{33}$. The relevant BPS equations are given
by
\begin{eqnarray}
\phi_1'&=&\frac{1}{8}e^{-\frac{\sigma}{2}-2\phi_1-\phi_2}(e^{4\phi_1}-1)\left[g_1+g_2+(g_1-g_2)e^{2\phi_2}\right],\label{Eq1}\\
\phi_2'&=&\frac{1}{8}e^{-\frac{\sigma}{2}-2\phi_1-\phi_2}\left[g_1(1+e^{2\phi_1})^2(e^{2\phi_2}-1)-g_2(1+e^{2\phi_2})
(e^{2\phi_1}-1)^2\right],\label{Eq2}\\
\sigma'&=&\frac{1}{20}e^{-\frac{\sigma}{2}-2\phi_1-\phi_2}\left[g_2(e^{2\phi_2}-1)(e^{2\phi_1}-1)^2-
g_1(1+e^{2\phi_1})^2(1+e^{2\phi_2})\phantom{e^{\frac{1}{2}}}\right.\nonumber \\
& &\left.-128he^{\frac{5\sigma}{2}+2\phi_1+\phi_2} \right],\label{Eq3}\\
A'&=&\frac{1}{40}e^{-\frac{\sigma}{2}-2\phi_1-\phi_2}\left[g_2(e^{2\phi_2}-1)(e^{2\phi_1}-1)^2-
g_1(1+e^{2\phi_1})^2(1+e^{2\phi_2})\phantom{e^{\frac{1}{2}}}\right.\nonumber \\
& &\left.+32he^{\frac{5\sigma}{2}+2\phi_1+\phi_2}
\right].\label{Eq4}
\end{eqnarray}
These equations reduce to the $SO(3)_{\textrm{diag}}$ case when
$\phi_2=\phi_1$. If we set $\phi_2=0$, consistency requires that
$\phi_1=0$. For $\phi_1=0$, the solution has $SO(2)_R\times SO(2)$
symmetry. This gives rise to the same solution studied above.
\\
\indent Since there are no interesting truncations, we now consider
a solution to the above equations with $\phi_1,\phi_2\neq 0$.
Finding the solution for a general value of $g_2$ turns out to be
difficult. However, for $g_2=g_1=-16h$, we can find an analytic
solution. The first step in finding this solution is to combine
\eqref{Eq1} and \eqref{Eq2} into a single equation
\begin{equation}
\frac{d\phi_2}{d\phi_1}=\frac{1+e^{4\phi_1}-2e^{2\phi_1+\phi_2}}{1-e^{4\phi_1}}
\end{equation}
which is solved by
\begin{equation}
\phi_2=\phi_1-\frac{1}{2}\ln
\left[\frac{8C_2-1+e^{4\phi_1}}{8C_2}\right].
\end{equation}
Changing to a new radial coordinate $\tilde{r}$ via
$\frac{d\tilde{r}}{dr}=e^{-\frac{\sigma}{2}-\phi_2}$, we obtain the
solution to equation \eqref{Eq1}
\begin{equation}
\phi_1=\pm\frac{1}{2}\ln \left[\frac{1+e^{C_1-16h
\tilde{r}}}{1-e^{C_1-16h \tilde{r}}}\right].
\end{equation}
To find the solution for $\sigma$, we change to another new
coordinate $R$ via
$\frac{dR}{dr}=-e^{-\frac{\sigma}{2}-\phi_2-2\phi_1}$. Equations
\eqref{Eq1},\eqref{Eq2} and \eqref{Eq3} can be combined to
\begin{equation}
\frac{5}{2}\frac{d\sigma}{dR}+2\frac{d\phi_1}{dR}+\frac{d\phi_2}{dR}=
-16h\left(1-e^{\frac{5}{2}\sigma+2\phi_1+\phi_2}\right)
\end{equation}
which gives
\begin{equation}
\sigma=-\frac{2}{5}\left[2\phi_1+\phi_2+\ln
\left(1-C_3e^{16hR}\right)\right].
\end{equation}
Combing \eqref{Eq3} and \eqref{Eq4}, we find an equation for $A$ as a function of $R$
\begin{equation}
\frac{dA}{dR}-\frac{1}{2}\frac{d\sigma}{dR}=-4e^{\frac{5}{2}\sigma+2\phi_1+\phi_2}
\end{equation}
whose solution, after using $\sigma$ solution, is given by
\begin{equation}
A=\frac{\sigma}{2}+\frac{1}{4}\ln
\left[C_3-e^{-16hR}\right].
\end{equation}
 As in the previous case, we separately consider the two
possibilities for $\phi_1>0$ and $\phi_1<0$.
\\
\indent For $\phi_1>0$, we can find the relation between $R$ and $\tilde{r}$ by using the relation $\frac{dR}{d\tilde{r}}=-e^{-2\phi_1(\tilde{r})}$. This results in
\begin{equation}
8hR=8h\tilde{r}-\ln\left[2(e^{C_1}+e^{16h\tilde{r}})\right].
\end{equation}
In term of $\tilde{r}$, the $\sigma$ and $A$ solutions become
\begin{eqnarray}
\sigma &=&-\frac{2}{5}\left[2\phi_1+\phi_2+\ln\left[1-\frac{C_3e^{16h\tilde{r}}}{4(e^{C_1}+e^{16h\tilde{r}})^2}
\right]\right],\label{sigma_sol_SO2_1}\\
A&=&\frac{\sigma}{2}+\frac{1}{4}\ln\left[C_3-4e^{-16h\tilde{r}}(e^{C_1}+e^{16h\tilde{r}})^2\right].\label{A_sol_SO2_1}
\end{eqnarray}
\indent Near the IR singularity at $16h\tilde{r}\sim C_1$, we have
$\phi_2\sim -\phi_1$ for all values of $C_2$. In the IR, the
solution behaves differently for $C_3=16e^{C_1}$ and $C_3\neq
16e^{C_1}$. This is because the logarithmic term in
\eqref{sigma_sol_SO2_1} and \eqref{A_sol_SO2_1} diverges, in this
limit, when $C_3=16e^{C_1}$. For $C_3\neq 16e^{C_1}$, we find
\begin{eqnarray}
& &\phi_1\sim-\phi_2\sim -\frac{1}{2}\ln (16h\tilde{r}-C_1),\qquad
\sigma \sim -\frac{2}{5}\phi_1\sim
\frac{1}{5}\ln (16h\tilde{r}-C_1),\nonumber \\
& &A\sim \frac{\sigma}{2}\sim \frac{1}{10}\ln (16h\tilde{r}-C_1),
\qquad ds^2=(16hr-C)^\frac{1}{8}dx^2_{1,5}+dr^2\, .
\end{eqnarray}
This gives rise to $V\rightarrow \infty$ which is physically
unacceptable.
\\
\indent However, if $C_3=16e^{C_1}$, the solution becomes
\begin{eqnarray}
\sigma &\sim &-\frac{3}{5}\ln (16h\tilde{r}-C_1),\qquad A\sim \frac{1}{5}\ln (1h\tilde{r}-C_1),\nonumber \\
ds^2 &=&(16hr-C)^{\frac{1}{3}}dx^2_{1,5}+dr^2\, .
\end{eqnarray}
This gives $V\rightarrow -\infty$, so this singularity is
acceptable. We see that flows with $\phi_1>0$ are physical provided
that $C_3=16e^{C_1}$.
\\
\indent For $\phi_1<0$, the solution $\phi_1=-\frac{1}{2}\ln \left[\frac{1+e^{C_1-16h\tilde{r}}}{1+e^{C_1-16h\tilde{r}}}\right]$ gives
\begin{equation}
8hR=8h\tilde{r}-\ln \left[2(e^{C_1}-e^{16h\tilde{r}})\right].
\end{equation}
Accordingly, the solutions for $\sigma$ and $A$ become
\begin{eqnarray}
\sigma &=&-\frac{2}{5}\left[2\phi_1+\phi_2+\ln\left[1-\frac{C_3e^{16h\tilde{r}}}{4(e^{C_1}-e^{16h\tilde{r}})^2}
\right]\right],\label{sigma_sol_SO2_2}\\
A&=&\frac{\sigma}{2}+\frac{1}{4}\ln\left[C_3-4e^{-16h\tilde{r}}(e^{C_1}-e^{16h\tilde{r}})^2\right].\label{A_sol_SO2_2}
\end{eqnarray}
In this case, the logarithmic term in \eqref{A_sol_SO2_2} diverges
as $16h\tilde{r}\sim C_1$ when $C_3=0$, but the logarithmic term in
\eqref{sigma_sol_SO2_2} vanishes. When $C_3\neq 0$, the situation is
reversed. Unlike the $\phi_1>0$ case, the value of $C_2$ is
important since there are two possibilities $\phi_1=\mp\phi_2$
depending $C_2=\frac{1}{8}$ or $C_2\neq\frac{1}{8}$.
\\
\indent We begin with the first case with $C_2=\frac{1}{8}$ and
$C_3=0$. The IR behavior of the solution is given by
\begin{eqnarray}
& &\phi_1\sim -\phi_2\sim \frac{1}{2}\ln (16h\tilde{r}-C_1),\qquad
\sigma \sim
\frac{1}{5}\ln (16h\tilde{r}-C_1),\nonumber \\
& &A\sim \frac{3}{5}\ln (16h\tilde{r}-C_1),\qquad
16hr-C=\frac{5}{3}(16h\tilde{r}-C_1)^{\frac{3}{5}}\,
.
\end{eqnarray}
The metric becomes
\begin{equation}
ds^2=(16hr-C)^{2}dx^2_{1,5}+dr^2\, .
\end{equation}
When $C_3\neq 0$, the solution in the IR becomes
\begin{eqnarray}
& &\phi_1\sim -\phi_2\sim \frac{1}{2}\ln (16h\tilde{r}-C_1),\qquad
\sigma \sim
\frac{3}{5}\ln (16h\tilde{r}-C_1),\nonumber \\
& &A\sim \frac{3}{10}\ln (16h\tilde{r}-C_1),\qquad
ds^2=(16hr-C)^{\frac{3}{4}}dx^{2}_{1,5}+dr^2\, .
\end{eqnarray}
Both of them lead to $V\rightarrow -\infty$.
Therefore, the solution with $\phi_1<0$ and $C_2=\frac{1}{8}$ is physical
for all values of $C_3$.
\\
\indent For $C_2\neq \frac{1}{8}$, we find, with $C_3=0$, the IR
behavior of the solution
\begin{eqnarray}
& &\phi_1\sim \phi_2\sim \frac{1}{2}\ln (16h\tilde{r}-C_1),\qquad
\sigma \sim
-\frac{6}{5}\ln (16h\tilde{r}-C_1),\nonumber \\
& &ds^2=(16hr-C)^{-\frac{2}{9}}dx^{2}_{1,5}+dr^2,
\end{eqnarray}
and, for $C_3\neq 0$,
\begin{eqnarray}
& &\phi_1\sim \phi_2\sim \frac{1}{2}\ln (16h\tilde{r}-C_1),\qquad
\sigma \sim
\frac{1}{5}\ln (16h\tilde{r}-C_1),\nonumber \\
& &ds^2=(16hr-C)^{\frac{1}{8}}dx^{2}_{1,5}+dr^2\, .
\end{eqnarray}
Both of them lead to $V\rightarrow \infty$. We then conclude that
flows with $\phi_1<0$ and $C_2\neq \frac{1}{8}$ are not physical for
any $C_3$.
\\
\indent It could be very interesting to have interpretations of
these results in terms of six-dimensional gauge theories.

\section{Conclusions}\label{conclusion}
We have studied some critical points of $N=2$, $SO(4)$ gauged
supergravity in seven dimensions. We have found one new
supersymmetric $AdS_7$ critical point with $SO(3)$ symmetry.
Recently, many new $AdS_7\times M_3$ solutions have been identified
in massive type IIA theory \cite{All_AdS7}. It would be interesting
to see weather the new supersymmetric $AdS_7$ obtained here could be
related to the classification in \cite{All_AdS7}. We have also found
a number of non-supersymmetric $AdS_7$ critical points and checked
their stability by computing all of the scalar masses. We have found
that although the non-supersymmetric critical point originally found
in pure gauged supergravity has been shown to be stable, it is
unstable in the presence of vector multiplet scalars. On the other
hand, new stable non-supersymmetric points are discovered here and
should correspond to new non-trivial IR fixed points of the $(1,0)$
SCFT.
\\
\indent An analytic RG flow solution interpolating between the
$SO(3)$ supersymmetric critical point and the trivial point with
$SO(4)$ symmetry has also been given. To the best of the author's
knowledge, this is the first example of holographic RG flows between
two supersymmetric fixed points of the $(1,0)$ field theory in six
dimensions. We have further studied supersymmetric flows to
non-conformal field theories and identified the physical flows.
These would provide more general flow solutions than those
considered in \cite{Ahn_7D_flow} and \cite{non_SUSY7Dflow} and could
be useful in a holographic study of the dynamics of six-dimensional
gauge theories similar to the analysis of \cite{confining_from_RG}.
Finding a field theory interpretation of the gravity solutions
obtained in this paper is also interesting.
\\
\indent We end the paper with a short comment on a more general
situation with $n$ vector multiplets. The $(1,0)$ field theory with
$E_8$ symmetry considered in \cite{Berkooz_6D_dual} would need
$n=248+3$ vector multiplets. The resulting gauge group in this case
is $SO(4)\times E_8$. The total $3\times (248+3)$ scalars, living on
$SO(3,248+3)/SO(3)\times SO(248+3)$ coset manifold, and the dilaton
transform as $(\mathbf{3},\mathbf{3},\mathbf{1})$,
$(\mathbf{3},\mathbf{1},\mathbf{248})$ and
$(\mathbf{1},\mathbf{1},\mathbf{1})$ under $SO(3)_R\times
SO(3)\times E_8$. We have considered only
$(\mathbf{3},\mathbf{3},\mathbf{1})$ and
$(\mathbf{1},\mathbf{1},\mathbf{1})$ scalars which are $E_8$
singlets. It is also interesting to consider scalars in
$(\mathbf{3},\mathbf{1},\mathbf{248})$ representation. Our solutions
given in this paper are of course solutions of the theory with
$SO(4)\times E_8$ gauge group by the group theory argument of
\cite{warner}.

\acknowledgments This work is partially supported by Chulalongkorn
University through Ratchadapisek Sompote Endowment Fund under grant
GDNS57-003-23-002 and The Thailand Research Fund (TRF) under grant
TRG5680010. The author gratefully thanks Chinnapat Panwisawas for
his help in finding old literatures.


\end{document}